\documentclass[twocolumn,doublespacing,showpacs,preprintnumbers,amsmath,amssymb]{revtex4}
\usepackage{amssymb}
\usepackage{txfonts}
\usepackage{graphicx}
\usepackage{subfigure}
\usepackage{dcolumn}
\usepackage{bm}
\usepackage{sidecap}

\begin{document}

\title{Passive scheme with photon-number-resolving detector for
monitoring the untrusted source in Plug-and-Play quantum key distribution system}
\author{Bingjie Xu}
\author{Xiang Peng}
\thanks{xiangpeng@pku.edu.cn.}
\author{Hong Guo}
\thanks{hongguo@pku.edu.cn.}
\affiliation{CREAM Group, State Key Laboratory of Advanced Optical Communication Systems and Networks (Peking University) and Institute of Quantum Electronics, School of Electronics Engineering and Computer Science, Peking University, Beijing 100871, PR China}
\date{\today}
\begin{abstract}
A passive scheme with a beam splitter and a photon-number-resolving (PNR)
detector is proposed to verify the photon statistics of an untrusted source in a plug-and-play
quantum-key-distribution system by applying a three-intensity decoy-state protocol. The practical
issues due to statistical fluctuation and detection noise
are analyzed. The simulation results show that the scheme can work
efficiently when the total number of optical pulses sent from Alice
to Bob is above $10^{8}$, and the dark count rate
of the PNR detector is below 0.5 counts/pulse, which is realizable with
current techniques. Furthermore, we propose a practical realization of the
PNR detector with a variable optical attenuator combined with a
threshold detector.
\end{abstract}
\maketitle

\section{Introduction}
Based on the BB84 protocol~\cite{BB84}, the security analysis for the practical
quantum-key-distribution (QKD) source was given~\cite{ILM_07,GLLP_04}.
Further, the decoy-state protocol was proposed~\cite{Decoy_Hwang_03} and was developed~\cite{Decoy_Lo_04,Decoy_Wang_05,Decoy_Lo_05,Decoy_Wang_PRA_05,Decoy_Ma_PRA_05,Mauerer_PRA_07} to improve
the QKD performance. Commonly, a trusted QKD source is considered for those
protocols, which means the photon-number distribution (PND) of the source is fixed
and is known by Alice and Bob. However, this assumption is
not always valid in practice. For example, the intensity fluctuation from the
source and the parameter fluctuation from the optical devices cause the assumption
of the trusted source to fail~\cite{UntruQKD_Wang_07}. In particular, an untrusted
source exists in a real-life experiment (i.e., two-way plug-and-play system)
and gives rise to the possibility of a Trojan-horse attack~\cite{TrojanAttack_06,
UntruQKD_Zhao_08,UntruQKD_Peng_08,UntruQKD_Zhao_09,UntruQKD_Peng_09}, where the
source is pessimistically controlled by Eve. Thus, the statistical characteristics
of the QKD source need to be verified to boost the QKD performance~\cite{TrojanAttack_06,
UntruQKD_Wang_PRA_07,UntruQKD_Wang_APL_07,UntruQKD_Wang_08,UntruQKD_Wang_09,UntruQKD_Wang_10,
UntruQKD_Zhao_08,UntruQKD_Zhao_09,UntruQKD_Peng_08,UntruQKD_Peng_09,UntruQKD_Moroder_09,
UntruQKD_Curty_09,UntrusQKD_Curty_10,UntrusQKD_Adachi_09,UntruQKD_Guo_09}.

Intuitively, if the characteristics of the untrusted source infinitely
approaches that of the trusted source, Alice needs a quantum nondemolition
(QND) measurement~\cite{Scully_97} to verify the PND of the QKD source. However,
it is hard to implement the QND measurement in practice. Fortunately, an analytical
method was provided to calculate the final key rate when the
probability of untagged bits was known by Alice and Bob~\cite{UntruQKD_Zhao_08},
and an active photon-number-analyzer (PNA) scheme
was proposed to monitor the probability of untagged bits.
However, it is challenging to implement the
active scheme, where a high speed random optical switch and
a perfect intensity monitor are needed. Then, using inverse-Bernoulli transformation, a passive scheme, with
a beam splitter (BS) and an imperfect detector, was proposed and was verified experimentally~\cite{UntruQKD_Peng_08}.
Furthermore, to realize the passive scheme
more efficiently, a high-speed two-threshold detection can be
used without applying inverse-Bernoulli transformation postprocessing~\cite{UntruQKD_Zhao_09,UntruQKD_Peng_09}.
From another viewpoint, recent results for the three-intensity decoy-state protocol with
the untrusted source have rigorously proved that it is sufficient to monitor the lower
and upper bounds of the probability for Alice to send out vacuum, one photon,
and two photon states~\cite{UntruQKD_Wang_08,UntruQKD_Wang_09}. More recently,
the detector-decoy scheme was theoretically proposed to monitor the PND of an
untrusted source using a threshold detector combined with a variable optical
attenuator (VOA)~\cite{UntruQKD_Moroder_09}.

In the following, a passive scheme with a BS and a
photon-number-resolving (PNR) detector, which can discriminate vacuum, one-photon,
two-photon, and more than two-photon states, is proposed to monitor the
parameters needed in Refs.~\cite{UntruQKD_Wang_08,UntruQKD_Wang_09}. Then,
the untrusted source in the plug-and-Play QKD system can be monitored with the passive scheme. Additionally, some practical
issues due to finite-data size and random-detection noise are included in the
analysis. Furthermore, a proposed realization of the PNR detector is analyzed
based on the idea of the detector-decoy scheme~\cite{UntruQKD_Moroder_09}.

\section{Key Parameters in security analysis}
Generally, the secure key rate of the BB84 protocol is~\cite{ILM_07,GLLP_04}
\begin{equation}\label{Eq:R BB84}
R=\frac{1}{2}Q\left\{\Delta _1[1-H_2(e_1)]-H_2(E)\right\},
\end{equation}
where $Q$ and $E$ are, respectively, the count rate and the quantum bit error rate (QBER) measured
in the QKD experiment, $\Delta _1$ ($e_1$) is the fraction of counts
(QBER) due to the single-photon state, and $H_2(x)=-x\log_2(x)-(1-x)\log_2(1-x)$
is the binary Shannon entropy. In the standard security analysis of the BB84 protocol,
all the losses and errors are assumed from the single-photon state \cite{GLLP_04},
which gives
\begin{equation}\label{Eq:Delta1_e1 BB84}
\begin{aligned}
\Delta _1=\frac{{Q - P_{multi} }}{Q},\  e_1=\frac{E}{\Delta_1},
\end{aligned}
\end{equation}
where $P_{multi}$ is the probability for Alice to send out multiphoton states.

The decoy-state method offers a more effective way to estimate the lower (upper)
bound of $\Delta _1$ ($e_1$) compared to Eq.~(\ref{Eq:Delta1_e1 BB84})~\cite{Decoy_Hwang_03,
Decoy_Lo_04,Decoy_Wang_05,Decoy_Lo_05,Decoy_Wang_PRA_05,Decoy_Ma_PRA_05,Mauerer_PRA_07}.
In the three-intensity decoy-state protocol \cite{{Decoy_Wang_05},Decoy_Ma_PRA_05},
Alice randomly sends three kinds of sources: vacuum, decoy, and signal sources,
respectively. The quantum state of the decoy (signal) source is $\rho_d=\sum_{n = 0}^\infty
{a_n \left| n \right\rangle }\left\langle n \right|$ ($ \rho _s  = \sum_{n = 0}^\infty
{a'_n \left| n \right\rangle } \left\langle n \right|$). For a three-intensity decoy-state
QKD system with an untrusted source, it was proved that~\cite{UntruQKD_Wang_08,UntruQKD_Wang_09}
\begin{equation}\label{Eq:Delta1 WXB}
\Delta _1^s  \ge \frac{{{a'_1}^{L} \left({a'_2}^{L} Q_d - a_2^U Q_s
- {a'_2}^{L} a_0^U Q_0  + a_2^U {a'_0}^{L} Q_0\right)}}{{Q_s
\left(a_1^U {a'_2}^{L}  - {a'_1}^{L} a_2^U \right)}},
\end{equation}
where $Q_0$, $Q_d$, or $Q_s$ is the count rate of vacuum, decoy,
and signal sources, respectively, and $\Delta _1^s$ is the fraction
of counts due to the single-photon state in the signal source. To calculate
the lower bound of $\Delta _1^s$, one needs to estimate the parameters
$\{$${a'_0}^L$, $a_0^U$, ${a'_1}^{L}$, $a_1^U$, ${a'_2}^{L}$,
$a_2^U\}$, where the superscript $L(U)$ means lower (upper) bound.
The secure key rate of the signal source is
\begin{equation} \label{Eq:key rate WXB}
R^s=\frac{1}{2}Q_s\{\Delta
_1^s[1-H_2(e_1^s)]-H_2(E_s)\},
\end{equation}
where $E_s$ is the QBER from the signal source and
$e_1^s=E_s/\Delta _1^s$. In the
following, we present a passive scheme to estimate the parameters
$\{$${a'_0}^L$, $a_0^U$, ${a'_1}^{L}$, $a_1^U$,
${a'_2}^{L}$, $a_2^U\}$.

\section{THEORY OF ESTIMATION with passive scheme}
\begin{figure}[b]
\begin{center}
\includegraphics[width=0.45\textwidth]{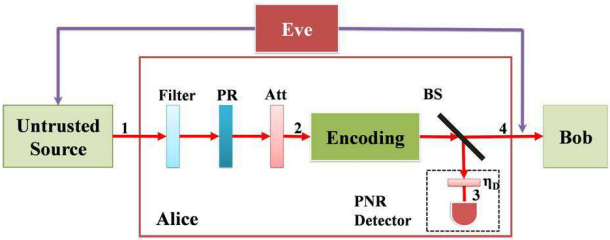}
\end{center}
\caption{(Color online) The untrusted source prepared at P$1$ by Eve, where P$i$ means position
 $i$ $(i=1,2,3,4)$, passes through an optical filter,
 a phase randomizer (PR), and an attenuator (Att) with the attenuation
 coefficient $\eta_s$ ($\eta_d$) for the signal (decoy) source. After the source
is encoded, a BS with transmittance $\eta_{BS}$ separates
 it into two beams: One goes to a PNR detector
with efficiency $\eta_D$ at P$3$, and the other is sent out from
Alice's side at P$4$. }\label{fig: exp scheme}
\end{figure}
The passive scheme for estimating the parameters $\{$${a'_0}^L$,
 $a_0^U$, ${a'_1}^{L}$, $a_1^U$, ${a'_2}^{L}$, $a_2^U\}$ is
shown in Fig.~\ref{fig: exp scheme}, where a PNR detector
that can discriminate the photon number of $n=0$, $n=1$, $n=2$, and $n\ge3$
is used. For simplicity, one can calibrate the setup to satisfy
\begin{equation}\label{eq:loss}
\eta_{D}(1-\eta_{BS})=\eta_{BS},
\end{equation}
where $\eta_{BS}$ is the transmittance of the BS and
$\eta_D$ is the detection efficiency of the PNR detector.
Under this condition, the PND at P$4$ is the same as
that at P$3$, where P$i$ means position $i$
$(i=1,2,3,4)$ in Fig.~\ref{fig: exp scheme}.

\begin{figure}[b]
\begin{center}
\includegraphics[width=0.5\textwidth]{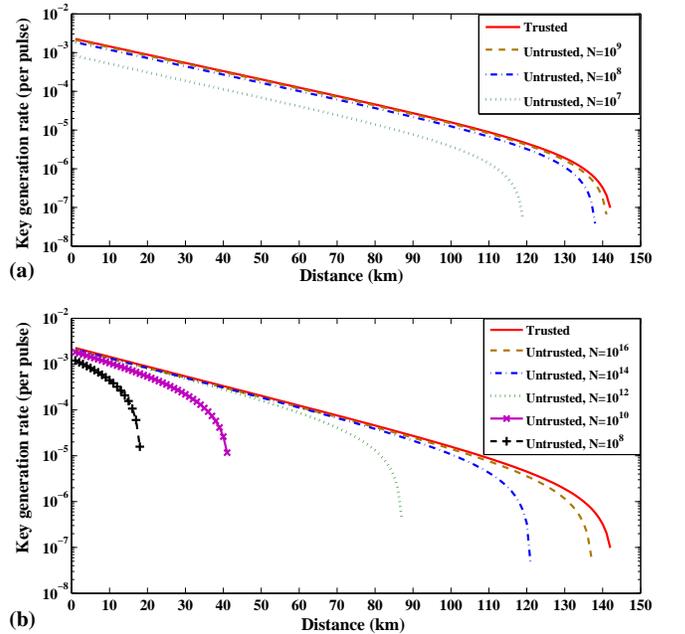}
\end{center}
\caption{(Color online) Simulation results of the three-intensity
decoy-state protocol for the trusted source with infinite-data size, and for the untrusted source with: (a) finite-data size $N ={10^9},\ {10^8},\ {10^7}$, respectively, based on the passive PNR
scheme in Fig.~1, where a BS and a noiseless PNR detector
are used to verify the parameters $\{$${a'_0}^L$,\ $a_0^U$,\ ${a'_1}^{L}$,
\ $a_1^U$,\ ${a'_2}^{L}$,\ $a_2^U\}$ with a confidence level
$1-10^{-6}$; (b) finite-data size $N
={10^{16}},\ {10^{14}},\ {10^{12}},\ {10^{10}},\ {10^9}$, respectively,
based on the passive PNA scheme \cite{UntruQKD_Zhao_09}, where a BS and noiseless PNA are used to verify the lower bound of
the probability of untagged bits with a confidence level
$1-10^{-6}$ (see Appendix A).}
\label{fig: finite data size}
\end{figure}

\subsection{PNR detector without detection noise}
In the following, the PNR detector is assumed to be noiseless.
Suppose that $P^{s(d)}(n_4)$ denotes the PND for the
signal (decoy) source at P4, and $D^{s(d)}(m)$ denote the PND
 for signal (decoy) source at P3. Clearly, one has
\begin{eqnarray}\label{Eq:a_n}
\nonumber a_n&=&P^{d}(n_4=n)=D^{d}(m=n),\\
a'_n&=&P^{s}(n_4=n)=D^{s}(m=n),
\end{eqnarray}
where $n=0,\ 1,\ 2,\cdots$.

Suppose that the data size $N$ is the total number of optical pulses sent from Alice to Bob, while
$N^{s(d)}$ is the number of signal (decoy) pulses, correspondingly.
Let $k_m^{s(d)}$ denote the number of detected signal (decoy) pulses at P3
given that the PNR detector records $m$ photoelectrons
$(m=0,1,2)$. Using the \emph{random sampling theory}
\cite{RandomSampling_37}, each $ D^s(m) \in [k_m^s /N^s- \varepsilon'
,k_m^s /N^s  + \varepsilon' ]$ with a confidence level $1-2\exp(-N^s
{\varepsilon'}^2/2)$ for signal pulses, and each $ D^d(m) \in [k_m^d /N^d
- \varepsilon ,k_m^d /N^d  + \varepsilon ]$ with a confidence level
$1-2\exp(-N^d \varepsilon^2/2)$ for decoy pulses can be estimated.
For enough small $\exp(-N^s
{\varepsilon'}^2/2)$ and $\exp(-N^d \varepsilon^2/2)$, simultaneously, $ D^s(m) \in [k_m^s /N^s- \varepsilon'
,k_m^s /N^s  + \varepsilon' ]$ and $ D^d(m) \in [k_m^d /N^d
- \varepsilon ,k_m^d /N^d  + \varepsilon ]$ for $m=0,1,2$ are estimated approximately with a confidence level $1-6\exp(-N^s {\varepsilon'}^2/2)-6\exp(-N^d {\varepsilon}^2/2)$.
From Eq.~(\ref{Eq:a_n}), one simultaneously gets
{\setlength\arraycolsep{2pt}
\begin{equation}\label{Eq: noiseless}
{a'}_{m}^{L}= \frac{{k_m^s }}{{N^s }} - \varepsilon',\ {a}_{m}^{U} =
\frac{{k_m^d }}{{N^d }} + \varepsilon,\ (m=0,1,2)
\end{equation}}
with a confidence level $1-6\exp(-N^s {\varepsilon'}^2/2)-6\exp(-N^d {\varepsilon}^2/2)$.

For testing the effects of the finite-data size, we choose an untrusted source of
Poissonian statistics to perform simulations based on the
three-intensity decoy-state protocol. Figure~\ref{fig: finite data
size}(a) shows the numerical simulation results for the trusted source,
and the untrusted source with the passive PNR scheme in Fig.~\ref{fig: exp
scheme}, where a BS and a noiseless PNR detector are used
to verify the parameters $\{$${a'_0}^L$, $a_0^U$, ${a'_1}^{L}$,
$a_1^U$, ${a'_2}^{L}$, $a_2^U\}$. Here, the average photon number
(APN) of the Poissonian source at P1 is $7.69 \times 10^{6}$. The
attenuation $\eta_{s(d)}$ is set to be $5\times10^{-7}(1\times10^{-7})$,
and the transmittance of the BS is $\eta_{BS}=0.13$ so that the APN for the signal (decoy)
state at P4 is $\mu_s=0.5$ ($\mu_d=0.1$). The detection efficiency
of the PNR detector $\eta_D$ is set to be 0.15, so that Eq.~(\ref{eq:loss}) holds. The photoelectron
detection data recorded by the PNR detector is simulated using the Monte Carlo
method, and $N =10^7, 10^8$, and $10^9$ of measurements are run.
Other experimental parameters are cited from the GYS experiment
\cite{GYS_04} as shown in Table~\ref{tab:para}, where $\eta_{Bob}$
is the efficiency of Bob's detection, $Y_0$ is the dark count rate of
Bob's detector, and $e_{\det}$ ($e_0$) is the probability that a
photon (dark count) hits the erroneous detector on Bob's side. To
compare the performance of the scheme in Fig.~\ref{fig: exp scheme}
with the passive PNA scheme proposed in Ref.~\cite{UntruQKD_Zhao_09}, where a
BS with transmittance $\eta_{BS}$ and a noiseless
PNA with efficiency $\eta_D$ are used to verify the probability of untagged bits (see Appendix A),
Fig.~2(b) shows the numerical simulation
for the trusted source and the untrusted source with the passive PNA scheme.
All the experimental parameters are chosen
to be the same as that for Fig.~2(a).

\subsection{PNR detector with additive detection noise}

Given a PNR detector with an independent
additive detection noise $y$, the detected photoelectron number $m'$,
and the photon number $m$ at P3 satisfy $m'=m+y$.
One can calculate the lower and upper bounds of PND
$D(m)\ (m=0,1,2)$ at P3 based on the detected photoelectron distribution $P(m')$
, given that the distribution of the detection
noise $N(y)$ is known by Alice and independent of the input source.
The dark count is the main kind of detection noise for the PNR
detectors, such as the time multiplexing detector
\cite{PNR_TMD_03,PNR_TMD_04}, the transition-edge sensor
\cite{PNR_TES_05}, or a threshold detector together with a variable
Att \cite{PNR_OnOff_05,UntruQKD_Moroder_09}. In the case of
independent Poissonian statistics noise, the
probability of detecting $m'$ photoelectrons is $P(m') = \sum^{m'}_{d=0}{N(d)} D(m'-d),$
where $N(y=d)= e^{ - \lambda }{\lambda ^d }/{d!}$ is the probability
that $d$ dark counts occur in the PNR detector, and $\lambda$ is
the average dark count rate. Then, one has
{\setlength\arraycolsep{1.5pt}\begin{equation}\label{Eq:Poissonian noise}
\left[ {\begin{array}{*{20}{c}}
   {D(m=0)}  \\
   {D(m=1)}  \\
   {D(m=2)}  \\
\end{array}} \right] = \left[ {\begin{array}{*{20}{c}}
   {P(m'=0)} & 0 & 0  \\
   {P(m'=1)} & {P(m'=0)} & 0  \\
   {P(m'=2)} & {P(m'=1)} & {P(m'=0)}  \\
\end{array}} \right]\left[ {\begin{array}{*{20}{c}}
   {{e^\lambda }}  \\
   { - {e^\lambda }\lambda }  \\
   {{e^\lambda }{\lambda ^2}/2}  \\
\end{array}} \right].
\end{equation}}
Let $k_{m'}^{s(d)}$ denote the number of detected signal (decoy)
pulses by Alice at P3 given that the PNR detector records $m'$
photoelectrons. Using the \emph{random-sampling theory}
\cite{RandomSampling_37}, simultaneously, $P^s(m') \in [k_{m'}^s /N^s  -
\varepsilon' ,k_{m'}^s /N^s  + \varepsilon' ]$ and
$ P^d(m') \in [k_{m'}^d /N^d - \varepsilon ,k_{m'}^d /N^d  +
\varepsilon ]$ for $m'=0,1,2$ are estimated with a confidence level
$1-6\exp(-N^s {\varepsilon'}^2/2)-6\exp(-N^d
\varepsilon^2/2)$. Combining Eqs.~(\ref{Eq:a_n}) and~(\ref{Eq:Poissonian noise}), one yields
{\setlength\arraycolsep{2pt}
\begin{eqnarray}\label{eq:an poisson noise}
 \nonumber a{'_0} &\ge& {e^\lambda }\left(\frac{{k_{m' = 0}^s}}{{{N^s}}} - \varepsilon' \right) = {a'_0}^{L}, \\
 \nonumber a{'_1} &\ge&  - \lambda {e^\lambda }\left(\frac{{k_{m' = 0}^s}}{{{N^s}}} + \varepsilon' \right) + {e^\lambda }\left(\frac{{k_{m' = 1}^s}}{{{N^s}}} - \varepsilon' \right) = {a'_1}^{L}, \\
 \nonumber a{'_2} &\ge& \frac{{{\lambda ^2}}}{2}{e^\lambda }\left(\frac{{k_{m' = 0}^s}}{{{N^s}}} - \varepsilon '\right) - \lambda {e^\lambda }\left(\frac{{k_{m' = 1}^s}}{{{N^s}}} + \varepsilon' \right) \\
 \nonumber & &{}+{e^\lambda }\left(\frac{{k_{m' = 2}^s}}{{{N^s}}} - \varepsilon' \right) = {a'_2}^{L},\\
  \nonumber {a_0} &\le& {e^\lambda }\left(\frac{{k_{m' = 0}^d}}{{{N^d}}} + \varepsilon \right) = a_0^U ,\\
  \nonumber {a_1} &\le&  - \lambda {e^\lambda }\left(\frac{{k_{m' = 0}^d}}{{{N^d}}} - \varepsilon \right) + {e^\lambda }\left(\frac{{k_{m' = 1}^d}}{{{N^d}}} + \varepsilon \right) = a_1^U ,\\
 \nonumber {a_2} &\le& \frac{{{\lambda ^2}}}{2}{e^\lambda }\left(\frac{{k_{m' = 0}^d}}{{{N^d}}} + \varepsilon \right) - \lambda {e^\lambda }\left(\frac{{k_{m' = 1}^d}}{{{N^d}}} - \varepsilon \right) \\
 & &{}+{e^\lambda }\left(\frac{{k_{m' = 2}^d}}{{{N^d}}} + \varepsilon \right) = a_2^U
\end{eqnarray}}
with a confidence level $1-6\exp(-N^s {\varepsilon'}^2/2)-6\exp(-N^d {\varepsilon}^2/2)$.
\begin{table}[hbt]
\caption{The simulation parameters for Figs.~\ref{fig: finite data size},~\ref{fig:detection noise} and~\ref{fig:sensitive to epsilon ZY}.}
\begin{ruledtabular}
\begin{tabular}{ccccccc}
$\eta_D$ &$\eta_{BS}$&$\eta_{Bob}$&$\alpha$&$Y_0$&$e_{\det}$&$e_0$\\
\hline $0.15$& 0.13&0.045&0.21&$1.7 \times 10^{-6}$&3.3\%&0.5
\end{tabular}
\end{ruledtabular}
\label{tab:para}
\end{table}

\begin{figure}[t]
\begin{center}
\includegraphics[width=0.50\textwidth]{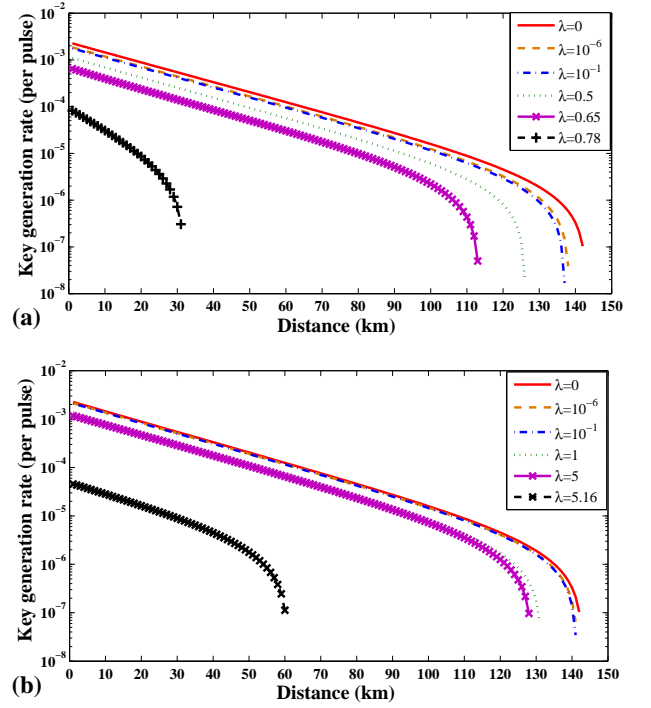}
\end{center}
\caption{(Color online) Simulation result of the three-intensity
decoy-state QKD with an untrusted source based on the scheme in Fig. 1:
(a)~The data size is $N=10^{8}$, and the APN of the
Poissonian noise in the PNR detector
is $\lambda=0$, $10^{-6}$, $10^{-1}$, $0.5$, $0.65$, $0.78$,
respectively; (b)~the data size is $N=10^{9}$, and the APN of the
Poissonian noise in the PNR detector
is $\lambda=0$, $10^{-6}$, $10^{-1}$, $1$, $5$, $5.16$,
respectively. The experimental parameters are the same as in Table I.
The confidence level for both cases is $1-10^{-6}$.}
\label{fig:detection noise}
\end{figure}

For testing the effects of detection noise, we choose an untrusted
source of Poissonian statistics to perform simulations based on the
three-intensity decoy-state protocol with the passive PNR scheme in
Fig.~\ref{fig: exp scheme}. The untrusted source is of Poissonian
statistics with APN $\mu=7.69\times 10^6$ at P1, and the attenuations
$\eta_s$ and $\eta_d$ are set to be $5\times10^{-7}$ and
$1\times10^{-7}$, respectively. The other experimental parameters
are cited from Table I. The photoelectron detection and additive
Poissonian noise of the PNR detector are simulated
using the Monte Carlo method, and $N= 10^8$ and $10^9$ of measurements
are run for Figs.~\ref{fig:detection noise}(a) and ~\ref{fig:detection noise}(b), respectively.
Our analysis is not limited to the Poissonian noise case. Generally,
when the random-positive detection noise $y$ with the probability $N(y)$ is
known to Alice, one has
{\setlength\arraycolsep{1.5pt}
\begin{equation}
\left[ {\begin{array}{*{20}{c}}
   {P(m' = 0)}  \\
   {P(m' = 1)}  \\
   {P(m' = 2)}  \\
\end{array}} \right] = \left[ {\begin{array}{*{20}{c}}
   {D(m = 0)} & 0 & 0  \\
   {D(m = 1)} & {D(m = 0)} & 0  \\
   {D(m = 2)} & {D(m = 1)} & {D(m = 0)}  \\
\end{array}} \right]\left[ {\begin{array}{*{20}{c}}
   {N(y = 0)}  \\
   {N(y = 1)}  \\
   {N(y = 2)}  \\
\end{array}} \right].
\end{equation}}

Thus, combining the results in Eqs.~(6) and~(10), one has
 {\setlength\arraycolsep{1pt}\begin{eqnarray}\label{eq:an general noise}
 \nonumber a{'_0} &\ge& \frac{{k_{m' = 0}^s/{N^s} - \varepsilon' }}{{N(y = 0)}}  ,\\
 \nonumber a{'_1} &\ge& \frac{{\left(k_{m' = 1}^s/{N^s} - \varepsilon' \right)N(y = 0) - \left(k_{m' = 0}^s/{N^s} + \varepsilon' \right)N(y = 1)}}{{{N^2}(y = 0)}} ,\\
 \nonumber a{'_2} &\ge& \frac{{k_{m' = 2}^s/{N^s} - \varepsilon' }}{{N(y = 0)}} - \frac{{k_{m' = 1}^s/{N^s} + \varepsilon' }}{{{N^2}(y = 0)}}N(y = 1)\\
 \nonumber &+&\frac{{k_{m' = 0}^s/{N^s} - \varepsilon '}}{{{N^3}(y = 0)}}{N^2}(y = 1) - \frac{{k_{m' = 0}^s/{N^s} + \varepsilon' }}{{{N^2}(y = 0)}}N(y = 2) ,\\
 \nonumber {a_0} &\le& \frac{{k_{m' = 0}^d/{N^d} + \varepsilon }}{{N(y = 0)}} ,\\
 \nonumber {a_1} &\le& \frac{{\left(k_{m' = 1}^d/{N^d} + \varepsilon \right)N(y = 0) - \left(k_{m' = 0}^d/{N^d} - \varepsilon \right)N(y = 1)}}{{{N^2}(y = 0)}} ,\\
 \nonumber {a_2} &\le& \frac{{k_{m' = 2}^d/{N^d} + \varepsilon }}{{N(y = 0)}} - \frac{{k_{m' = 1}^d/{N^d} - \varepsilon }}{{{N^2}(y = 0)}}N(y = 1) \\
 &+& \frac{{k_{m' = 0}^d/{N^d} + \varepsilon }}{{{N^3}(y = 0)}}{N^2}(y = 1) - \frac{{k_{m' = 0}^d/{N^d} - \varepsilon }}{{{N^2}(y = 0)}}N(y = 2).
 \end{eqnarray}}
Therefore, once the distribution of the detection noise is known, the
secure key rate can be estimated given the bounds of $\{$${a'_0}^L$,
 $a_0^U$, ${a'_1}^{L}$, $a_1^U$, ${a'_2}^{L}$, $a_2^U\}$.
\section{A proposed realization of a PNR detector}
\begin{figure}[htbp]
\begin{center}
\includegraphics[width=0.5\textwidth]{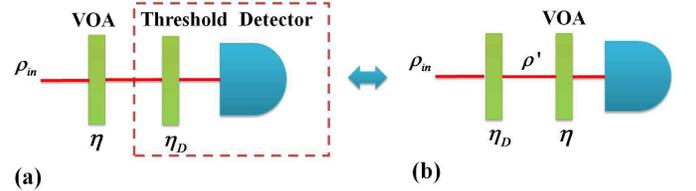}
\end{center}
\caption{(Color online) (a) A threshold detector (modeled by an
Att with transmittance $\eta_D$ and an ideal threshold
detector) combined with a VOA
(transmittance: $\eta$) can realize a PNR detector
\cite{UntruQKD_Moroder_09}. (b) An equivalent model to (a), which
means the two models will produce the same output
 given the same input. }\label{fig: PNR}
\end{figure}
The PNR detector can be realized by a VOA
(VOA) combined with a practical threshold detector  as shown in Fig.~4(a)
\cite{UntruQKD_Moroder_09}, which is equivalent to the model in
Fig.~4(b). Suppose that the state of input source is
$\rho_{in}=\sum^{\infty}_{n=0} p_n\left| n \right\rangle
\left\langle n \right|$. In figure~4(b), after passing
through an Att with efficient $\eta_D$, the state of the
source becomes
$
\rho'=\sum^{\infty}_{n=0}
p'_n\left| n \right\rangle \left\langle n \right|,
$
where $p'_n=\sum^{\infty}_{m=n}p_m\left( {\begin{array}{*{20}{c}}
   m  \\
   n  \\
\end{array}} \right)\eta_D^n(1-\eta_D)^{m-n}$. When Eq.~(\ref{eq:loss}) holds,
one has
\begin{equation}\label{eq:PNR}
{p'_n}^{s}=a'_n, \ {p'_n}^d=a_n.
\end{equation}
Then, the source passes through the VOA with efficiency $\eta$, and the probability that
the detector dose not click can be calculated as $
p(\eta)=\sum_{n=0}^\infty (1-\eta)^n p'_n$ \cite{UntruQKD_Moroder_09}.
When we take the dark count of the threshold detector into account,
it can be calculated as
$p(\eta)=(1-\lambda)\sum_{n=0}^\infty (1-\eta)^n p'_n$,
where $\lambda$ is the dark count rate of the detector.
If Alice varies the transmittance of the VOA $\eta\in\{\eta_1, \cdots,\eta_M\}$, she
has a set of linear equations,
\begin{equation}\label{eq:lineareqn}
 p(\eta_i)=(1-\lambda)\sum_{n=0}^\infty (1-\eta_i)^n p'_n,\ (i=1,\cdots,M).
 \end{equation}
When she employs an infinite number of possible transmittance $\eta\in[0,\ 1]$,
she can always estimate any finite number of probabilities $p'_n$ with arbitrary
precision by solving Eqs.~(\ref{eq:lineareqn}). However, it is not necessary
for our purpose in which we are mainly concerned with the  probability of vacuum, one-photon,
and two-photon states, and, thus, only three  different transmittances $\eta\in\{\eta_0,
\ \eta_1,\ \eta_2\}$ are needed~\cite{UntruQKD_Moroder_09}.
One can choose $\eta_0=1$,
\begin{equation}\label{eq:p0}
p(\eta_0=1)=(1-\lambda) p'_0.
\end{equation}
Then one has
{\setlength\arraycolsep{1pt}\begin{eqnarray}
\nonumber \frac{p(\eta_1)}{1-\lambda}&\ge& p'_0+(1-\eta_1)p'_1,\\
\nonumber \frac{p(\eta_1)}{1-\lambda}&\le& p'_0+(1-\eta_1)p'_1+(1-\eta_1)^2(1-p'_0-p'_1),
\end{eqnarray}}
from which one gets
\begin{widetext}
\begin{eqnarray}\label{eq:p1}
 p'_1\le \frac{p(\eta_1)-p(\eta_0)}{(1-\lambda)(1-\eta_1)}=\overline{p'_1} , \ p'_1\ge\frac{p(\eta_1)-p(\eta_0)[1-(1-\eta_1)^2]-(1-\lambda)(1-\eta_1)^2}{(1-\lambda)[1-\eta_1-(1-\eta_1)^2]}=\underline{p'_1}.
\end{eqnarray}
In a similar way, one has
{\setlength\arraycolsep{1pt}\begin{eqnarray}
\nonumber \frac{p(\eta_2)}{1-\lambda}&\ge&p'_0+(1-\eta_2)p'_1+(1-\eta_2)^2p'_2 \ge p'_0+(1-\eta_2)\underline{p'_1}+(1-\eta_2)^2p'_2\\
\nonumber \frac{p(\eta_2)}{1-\lambda} &\le& p'_0+(1-\eta_2)p'_1+(1-\eta_2)^2p'_2 +(1-\eta_2)^3(p'_3+p'_4+p'_5+\cdots)\\
\nonumber &\le&[1-(1-\eta_2)^3]p'_0+[1-\eta_2-(1-\eta_2)^3]\overline{p'_1}+[(1-\eta_2)^2-(1-\eta_2)^3]p'_2+(1-\eta_2)^3,
\end{eqnarray}}
from which the upper and lower bounds for $p'_2$ can be estimated as
{\setlength\arraycolsep{1pt}\begin{eqnarray}\label{eq:p2}
\nonumber p'_2&\le&\frac{p(\eta_2)-p(\eta_0)-(1-\lambda)(1-\eta_2)\underline{p'_1}}{(1-\lambda)(1-\eta_2)^2}=\overline{p'_2}\\
 p'_2&\ge& \frac{{{p({\eta _2})} - [1 - {{(1 - {\eta _2})}^3}]{p({\eta _0})} - (1 - \lambda )[1 - {\eta _2} - {{(1 - {\eta _2})}^3}]\overline {p{'_1}}  - (1 - \lambda ){{(1 - {\eta _2})}^3}}}{(1 - \lambda)[{{{(1 - {\eta _2})}^2} - {{(1 - {\eta _2})}^3}}]}
=\underline{p'_2}.
\end{eqnarray}}
\end{widetext}
In conclusion, based on the recorded data $\{p(\eta_0)$, $p(\eta_1)$, $p(\eta_2)\}$,
Alice can estimated the parameters $\{$${a'_0}^L$,
 $a_0^U$, ${a'_1}^{L}$, $a_1^U$, ${a'_2}^{L}$, $a_2^U\}$ as in Eqs.~(\ref{eq:PNR}) and~(\ref{eq:p0})-(\ref{eq:p2}).
The scheme in Fig.~\ref{fig: PNR} can easily be realized with
current technology. As for the effect of statistical fluctuation, one can use the \emph{random-sampling theory} as before to consider the fluctuation of the  $\{p(\eta_0),\ p(\eta_1),\ p(\eta_2)\}$
with a confidence level so that we still can bound $\{$${a'_0}^L$,
 $a_0^U$, ${a'_1}^{L}$, $a_1^U$, ${a'_2}^{L}$, $a_2^U\}$.

\section{discussion and conclusion}

The results in Fig.~\ref{fig: finite data size} show that: (i)~The
performance of a QKD system with an untrusted source is close to
that of a trusted source, when the source is monitored efficiently
and the data size is large enough; (ii)~finite-data size has negative
effect on the secure key rate; (iii)~the method in Ref.~\cite{UntruQKD_Zhao_09}
is more sensitive to statistical fluctuation and needs a larger data
size than the method proposed in this paper.

In the passive PNA scheme proposed in Ref.~\cite{UntruQKD_Zhao_09} (see
Fig.~\ref{fig:ZY scheme} in Appendix A), Alice uses a PNA
to monitor the probability of the untagged bits in the untrusted
source, after which, one can estimate the lower bound of the secure key rate
with a confidence level as shown in Eq.~(\ref{eq:ZY key rate}). When the
confidence level is set to be constant (e.g., $1-10^{-6}$), the
estimation resolution $\varepsilon$ for the probability of
untagged bits is only decided by the data size $N$ (ignoring the
effect of  detection noise), where the confidence level is
$1-2\exp(-N\varepsilon^2/4)$~\cite{UntruQKD_Zhao_09}. However, the
secure key rate in Ref.~\cite{UntruQKD_Zhao_09} is sensitive to the
estimation resolution $\varepsilon$, and will reduce greatly when
$\varepsilon$ increases (see Fig.~\ref{fig:sensitive to epsilon ZY}
in Appendix A). When the data size $N$ decreases, the resolution
$\varepsilon$ has to increase to keep the constant confidence level,
and, thus, the key rate will reduce.

While in the scheme shown in Fig.~\ref{fig: exp scheme}, Alice uses a PNR
detector to monitor the probability of vacuum, one-photon, and
two-photon states for the signal and decoy sources, respectively. Because
of the low intensity of the output pulses at P4 (e.g., $\mu_s=0.5,\
\mu_d=0.1$), the vacuum, one-photon, and two-photon pulses are dominant
in pulses, and Alice can gain most of the information
about the statistics of the untrusted source at P4 based on the
recorded data of the PNR detector. In our scheme, six parameters are monitored,
and more information is gained than from
the scheme in Ref.~\cite{UntruQKD_Zhao_09}. Mathematically, the formulas shown by
Eq.~(\ref{Eq:key rate WXB}) are not so sensitive to the
estimation resolution of $\{$${a'_0}^L$, $a_0^U$, ${a'_1}^{L}$,
$a_1^U$, ${a'_2}^{L}$, $a_2^U\}$ compared to that in Ref.~
\cite{UntruQKD_Zhao_09} so that it does not require a very large
data size to work efficiently, as shown in Fig.~2(a). When the data
size is $N\ge10^8$, the performance of the scheme is very close to that
of the trusted source. In the asymptotic case where Alice sends infinitely
long bits of sequence ($N \sim \infty$), the performance will be the same as that of a trusted source as shown in Appendix B.

The results in Fig.~\ref{fig:detection noise} show that: (i)~Given
a PNR detector with the same dark count rate, the performance of a
system with an untrusted source will be better when the data size increases;
(ii)~given the same data size, the performance of a system with an untrusted
source will reduce when the dark count rate increases. The performance of the
scheme in Fig.~\ref{fig: exp scheme} is quite sensitive to the detection
noise of the PNR detector. It is shown that when the data size is $N\ge10^8$
and the dark count rate of the PNR detector is $\lambda\le 0.5$ counts/pulse, which are
realizable by current techniques \cite{PNR_TMD_03,PNR_TMD_04,PNR_TES_05,PNR_OnOff_05},
this scheme can still work efficiently.

In conclusion, we propose an experimental scheme to verify the key
parameters needed in Refs.~\cite{UntruQKD_Wang_08,UntruQKD_Wang_09}. The practical issues due to
detection noise and finite-data-size fluctuation are analyzed. We
also propose a realization of the PNR detector based on the detector-decoy
method, which is very practical in real experiment. Therefore,
the passive scheme with a PNR detector is highly practical to solve the untrusted source problem
in the two-way plug-and-play QKD system. This passive scheme is also applicable to
monitor the intensity fluctuation in a one-way QKD system, where an active scheme has
been proposed and has been tested experimentally~\cite{UntruQKD_Guo_09}.

We remark that the effect of parameter fluctuations has not
yet been included in the security analysis. The effective method
to deal with the parameter fluctuations [19] is encouraged
to be applied in the passive scheme.

\begin{acknowledgments}
This work was supported by the Key Project of National Natural
Science Foundation of China (Grant No. 60837004) and the National
Hi-Tech Research and Development (863) Program. X. Peng acknowledges
financial support from the China Postdoctoral Science Foundation (Grant No. 20100470134).
\end{acknowledgments}

\appendix
\section{Passive scheme method in \cite{UntruQKD_Zhao_09}}
\begin{figure}[htbp]
\begin{center}
\includegraphics[width=0.4\textwidth]{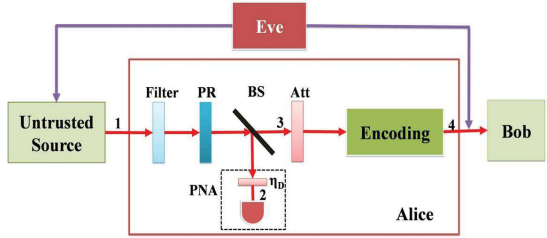}
\end{center}
\caption{(Color online) The model of the passive PNA scheme in Ref.~\cite{UntruQKD_Peng_08,UntruQKD_Zhao_09,UntruQKD_Peng_09}. The
untrusted source prepared by Eve passes through an
optical filter and a PR. Then, a BS with transmittance $\eta_{BS}$ separates it into two
beams: One goes to a PNA with efficiency $\eta_D$, and the other is attenuated by an Att with
efficiency $\eta_{s(d)}$ for the signal (decoy) state and is encoded before being from
sent out of Alice's side. }\label{fig:ZY scheme}
\end{figure}
\begin{figure}[htbp]
\begin{center}
\includegraphics[width=0.50\textwidth]{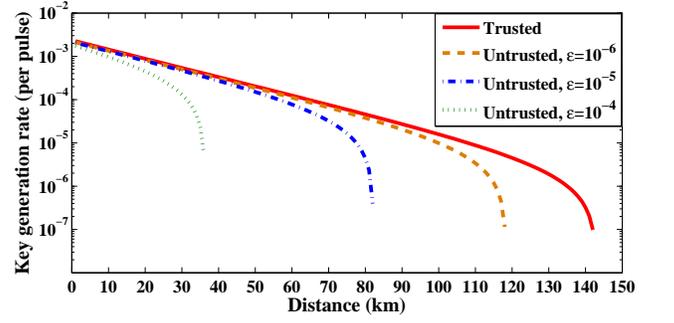}
\end{center}
\caption{(Color online) Simulation result of the three-intensity decoy
state protocol for the trusted source and the untrusted source with different estimation
resolutions $\varepsilon = 10^{-6},\ 10^{-5},\ 10^{-4}$, respectively. Based
on the passive PNA scheme in Fig.~\ref{fig:ZY scheme}, a BS and a noiseless PNA are used
to verify the lower bound of the probability of ¡°untagged bits¡±. Experimental
parameters are cited from Table~\ref{tab:para}.}\label{fig:sensitive to epsilon ZY}
\end{figure}

The passive PNA scheme in Ref.~\cite{UntruQKD_Zhao_09} is shown in Fig.~\ref{fig:ZY scheme}.
For simplicity, one can set $\eta_{D}(1-\eta_{BS})=\eta_{BS}$ so that
the PND at position 2 is the same as that at position 3 in Fig.~\ref{fig:ZY scheme}.
Define the pulses, whose photon number is $M \in [{M_{\min }},{M_{\max }}]$ at
position 3, as untagged bits. A BS and a PNA are used to
record the frequency of untagged bits experimentally.

Suppose that $N$ pulses are sent from Alice to Bob. Alice and Bob do
not know which bits are untagged bits. Let $N_{untagged}$ denote the
number of detected pulses by the PNA when the recorded
photoelectron number at position~2 belongs to $[{M_{\min }},{M_{\max
}}]$, and $\Delta  = {N_{untagged}}/N$. From the recorded data in
the PNA, one can estimate that at least $(1-\Delta-\varepsilon)N$ pulses
are untagged bits with a confidence $1-2\exp(-N\varepsilon^2/4)$
where $\varepsilon$ is a small positive parameter
\cite{UntruQKD_Zhao_09}.

Alice can measure the overall gain
$Q_{s(d)}$ and the QBER $E_{s(d)}$ for signal (decoy) pulses,
respectively, while she does not know the gain and the QBER for
the untagged bits. The upper and lower bounds of the gain of
the untagged bits for the signal (decoy) source can be estimated as
 \begin{equation*}\label{eq:gain of untagged}
\overline {Q_{s(d)}}= \frac{{{Q_{s(d)}}}}{{1 - \Delta  - \varepsilon
}}, \ \underline {Q_{s(d)}} = \max\left \{ 0,\frac{{{Q_{s(d)}} -
\Delta - \varepsilon }}{{1 - \Delta  - \varepsilon }}\right\}.
 \end{equation*}
The upper and lower bounds for the QBER of the untagged bits can be estimated as
 \begin{equation*}\label{eq:QBER of untagged}
 \overline {{Q_s}{E_s}}  = \frac{{{Q_s}{E_s}}}{{1 - \Delta  - \varepsilon }},\ \underline {{Q_s}{E_s}}  =\max\left\{0, \frac{{{Q_s}{E_s} - \Delta  - \varepsilon }}{{1 - \Delta  - \varepsilon }}\right\},
 \end{equation*}
for signal states, and
  \begin{equation*}\label{eq:QBER of untagged}
 \overline {{Q_d}{E_d}}  = \frac{{{Q_d}{E_d}}}{{1 - \Delta  - \varepsilon }},\ \underline {{Q_d}{E_d}}  =\max\left\{0, \frac{{{Q_d}{E_d} - \Delta  - \varepsilon }}{{1 - \Delta  - \varepsilon }}\right\},
 \end{equation*}
for decoy states. For the untagged bits, one can show that the upper and lower bounds
of the probability that the output photon number at position 4
is $n$ for signal (decoy) pulses are
\begin{equation*}\label{eq:mg1}
\overline {P_n^{s(d)}}=\left\{ \begin{array}{l@{\quad\quad } l}
                {{(1 - {\eta _{s(d)}})}^{{M_{\min }}}} & n=0, \\
                {\left( {\begin{array}{*{20}{c}}
   {{M_{\max }}}  \\
   n  \\
\end{array}} \right)\eta _{s(d)}^n{{(1 - {\eta _{s(d)}})}^{{M_{\max }} - n}}}
 & 1\le n \le M_{\max},\\
 0 &n>M_{\max},
 \end{array}
\right.
\end{equation*}
\begin{equation*}\label{mg2}
\underline {P_n^{s(d)}}=\left\{ \begin{array}{l@{\quad\quad } l}
                {{\left(1 - {\eta _{s(d)}}\right)}^{{M_{\max }}}} & n=0, \\
                {\left( {\begin{array}{*{20}{c}}
   {{M_{\min }}}  \\
   n  \\
\end{array}} \right)\eta _{s(d)}^n{{(1 - {\eta _{s(d)}})}^{{M_{\min }} - n}}}
 &1\le n \le M_{\min},\\
 0  & n>M_{\min},
 \end{array}
\right.
\end{equation*}
under the condition $M_{\max}\eta_{s(d)}<1$.

When the lower bounds of the probability of the untagged
bits are known by Alice, the secure key rate for the three-intensity
decoy-state protocol with an untrusted source is \cite{UntruQKD_Zhao_08}
\begin{equation}\label{eq:ZY key rate}
R = \frac{1}{2}\left\{  - {Q_s}{H_2}({E_s}) + (1 - \Delta  -
\varepsilon )\underline {Q_1^s} [1 - {H_2}(\overline {e_1^s}
)]\right\},
 \end{equation}
where
 \begin{eqnarray}
 \begin{aligned}
\nonumber \underline {Q_1^s}  =& \frac{{\underline {P_1^s}
}}{{\overline {P_1^d} \underline {P_2^s}  - \underline {P_1^s}
\overline {P_2^d} }}\times \left\{ \underline {{Q_d}} \underline
{P_2^s} -\overline {{Q_s}} \overline {P_2^d}  + \underline {P_0^s}
\overline {P_2^d} {Q_0} -\right.\\& \left.\overline {P_0^d}
\underline {P_2^s} {Q_0}-\frac{{({M_{\max }} - {M_{\min }}){{(1 -
{\eta _d})}^{{M_{\max }} - {M_{\min }} - 1}}\underline {P_2^s}
}}{{[{M_{\min }} + 1]!}}\right\},
\end{aligned}
\end{eqnarray}
 and
 \begin{eqnarray}
\nonumber \overline {e_1^s}  = \frac{{\overline {{E_s}{Q_s}}  - \underline {P_0^s} \underline {{E_0}{Q_0}} }}{{\underline {Q_1^s} }}.
 \end{eqnarray}
For testing the effects of $\varepsilon$ onto the secure key rate,
we choose an untrusted source of Poissonian statistics to perform
the simulations based on the three-intensity decoy-state protocol. Suppose that
the untrusted source has Poissonian PND with an APN of $7.69\times10^{6}$ at
position 1 of Fig.~5. Set $\eta_s=5\times10^{-7}$, and $\eta_d=1\times10^{-7}$.
The other experimental parameters are chosen to be the same as in Table I. The
values of $M_{\max}$ and $M_{\min}$ are chosen to be constant. The results
in Fig.~6 show that the final key rate is very sensitive to the value of $\varepsilon$.

Suppose that Alice has a noiseless PNA, and the
estimation confidence level is set to be constant
 \begin{equation}
 1-2e^{-N\varepsilon^2/4}=1-10^{-6}.
 \end{equation}
The estimation resolution
$\varepsilon$ is only decided by the data size $N$.
When the data size
$N$ decreases, the resolution $\varepsilon$ has to increase to keep
the constant confidence level, and, thus, the key rate will reduce
greatly as shown in Fig.~~2(b).

\section{Asymptotic case of method in \cite{UntruQKD_Wang_08}}
In the asymptotic case, Alice sends the infinitely long bits sequence ($N  \sim  \infty$).
Therefore, one can consider $\varepsilon,\ \varepsilon'\sim 0$ in
Eqs.~(\ref{Eq: noiseless}),~(\ref{eq:an poisson noise}), or~(\ref{eq:an general noise})
while still having the confidence level 1. Suppose that the PND of the untrusted source is Poissonian
with an APN of $\mu_{s(d)}$ for the signal (decoy) source at P4 in~Fig.\ref{fig: exp scheme}. One has
{\setlength\arraycolsep{1pt}\begin{eqnarray}
\nonumber &a_0^U& = e^{-\mu _d},\ a_1^U = {\mu _d}e^{ - {\mu _d}},\ a_2^U = \frac{\mu _d^2}{2}e^{- {\mu _d}},\\
\nonumber &{a'_0}^L&=e^{- {\mu _s}},\ {a'_1}^L = {\mu _s}e^{ - {\mu _s}},\ {a'_2}^L =\frac{ \mu _s^2}{2}e^{ - {\mu _s}}.
\end{eqnarray}}
Then, one can estimate
{\setlength\arraycolsep{1pt}
\begin{eqnarray}
\nonumber Q _1^s=Q_s \Delta _1^s &\ge& \frac{{{a'_1}^{L}
\left({a'_2}^{L} Q_d - a_2^U Q_s - {a'_2}^{L} a_0^U Q_0  + a_2^U
{a'_0}^{L} Q_0 \right)}}{{a_1^U {a'_2}^{L}  -
{a'_1}^{L} a_2^U }}\\
\nonumber&=&\frac{{{\mu _s}}}{{\mu _s^2 - {\mu _s}{\mu
_d}}}\left({Q_d}{e^{ - {\mu _d}}} - {Q_s}{e^{ - {\mu _s}}}\frac{{\mu
_d^2}}{{\mu _s^2}} - \frac{{\mu _s^2 - \mu _d^2}}{{\mu
_s^2}}{Q_0}\right),
\end{eqnarray}}
which is exactly the same as the case for a trusted source.

\end{document}